\documentclass[preprint,superscriptaddress,showpacs,preprintnumbers,amsmath,amssymb]{revtex4}

\usepackage{graphicx}% Include figure files
\usepackage{dcolumn}% Align table columns on decimal point
\usepackage{bm}% bold math
\usepackage{mathrsfs}
\usepackage{color}

\newcommand{\lb}{\left(}
\newcommand{\rb}{\right)}
\newcommand{\ls}{\left[}
\newcommand{\rs}{\right]}
\newcommand{\Lb}{\left\{}
\newcommand{\Rb}{\right\}}

\newcommand{\ff}[1]{\frac{1}{#1}}
\newcommand{\bo}{\boldsymbol}
\newcommand{\scr}[1]{{\mathscr #1}}

\newcommand{\mCB}{{\rm CB}}
\newcommand{\mSOP}{{\rm SOP}}

\begin{document}

%\preprint{APS/123-QED}

\title{Spin symmetry in Dirac negative energy spectrum
    in density-dependent relativistic Hartree-Fock theory}% Force line breaks with \\

\author{Haozhao Liang}%
 \affiliation{State Key Lab Nucl. Phys. {\rm\&} Tech., School of Physics, Peking University, Beijing 100871, China}
 \affiliation{Institut de Physique Nucl\'eaire, IN2P3-CNRS and Universit\'e Paris-Sud,
    F-91406 Orsay Cedex, France}

\author{Wen Hui Long}
 \affiliation{State Key Lab Nucl. Phys. {\rm\&} Tech., School of Physics, Peking University, Beijing 100871, China}
 \affiliation{Physik-Department der Technischen Universit\"at M\"unchen, D-85748 Garching, Germany}

\author{Jie Meng}
 \affiliation{School of Physics and Nuclear Energy, Beihang University,
              Beijing 100191, China}
 \affiliation{State Key Lab Nucl. Phys. {\rm\&} Tech., School of Physics, Peking University, Beijing 100871, China}
 \affiliation{Department of Physics, University of Stellenbosch, Stellenbosch, South Africa}

\author{Nguyen Van Giai}
 \affiliation{Institut de Physique Nucl\'eaire, IN2P3-CNRS and Universit\'e Paris-Sud,
    F-91406 Orsay Cedex, France}

\date{\today}

\begin{abstract}

The spin symmetry in the Dirac negative energy spectrum and its
origin are investigated for the first time within the
density-dependent relativistic Hartree-Fock (DDRHF) theory. Taking
the nucleus $^{16}$O as an example, the spin symmetry in the
negative energy spectrum is found to be a good approximation and the
dominant components of the Dirac wave functions for the spin
doublets are nearly identical. In comparison with the relativistic
Hartree approximation where the origin of spin symmetry lies in the
equality of the scalar and vector potentials, in DDRHF the
cancellation between the Hartree and Fock terms is responsible for
the better spin symmetry properties and determines the subtle
spin-orbit splitting. These conclusions hold even in the case when
significant deviations from the G-parity values of the
meson-antinucleon couplings occur.

\end{abstract}


\maketitle

\section{Introduction}

The relativistic Hartree approximation or relativistic mean field
(RMF) theory \cite{Serot:1986} has received much attention due to
its successful description of infinite nuclear matter as well as
finite nuclei near and far away from the $\beta$ stability line
\cite{Ring:1996,Vretenar:2005,Meng:2006}. One of its great success
is the natural description of the nuclear spin-orbit potential,
which leads to a remarkable spin-orbit splitting for the states with
the same orbital angular momentum and opposite spins $(j= l \pm
1/2)$, allowing for the understanding of the magic numbers and
forming the basis of nuclear shell structure. Furthermore, the
pseudo-spin symmetry \cite{Arima:1969, Hecht:1969}, i.e., the near
degeneracy between two single-particle states with the quantum
numbers $(n, l, j = l + 1/2)$ and $(n-1,l+2,j=l+3/2)$, whose origin
was a long mystery in nuclear physics \cite{Bohr:1982,Bahri:1992},
is well interpreted within the relativistic scheme with local
potentials (see Ref. \cite{Ginocchio:2005} and references therein).
The conservation and realization of pseudo-spin symmetry were
discussed in detail within the RMF framework \cite{Meng:1998PRCps,
Meng:1999,Marcos:2000, Marcos:2001,Chen:2003}. With the same origin,
the spin symmetry in the Dirac negative energy spectrum (i.e. the
single anti-nucleon spectrum) was proposed and investigated in RMF
theory \cite{Zhou:2003, He:2006}.

As the Fock terms are missing and the one-pion exchange potential is
not explicitly included in RMF, for the completeness of the theory,
there have been attempts to include the Fock terms in the
ground-state energy of nuclear systems over the past two decades
\cite{Bouyssy:1985, Bouyssy:1987, Bernardos:1993,Marcos:2004}.
Recently, the RHF theory with density-dependent nucleon-meson
couplings (DDRHF) finally succeeded in the quantitative description
of the ground-state properties of many nuclear systems on the same
level as RMF \cite{Long:2006}. Furthermore, it is found that the
DDRHF theory can improve the descriptions of the nucleon effective
mass and its isospin and energy dependences \cite{Long:2006}, as
well as the shell evolution and closure with the inclusion of the
one-pion exchange and $\rho$-tensor correlations \cite{Long:2008,
Long:2007}. The pseudo-spin symmetry and its origin as well as the
importance of the Fock terms have also been investigated before
\cite{Lopez:2003, Long:2006PS}. Although the pseudo-spin symmetry is
still found to be a good approximation in RHF, its mechanism becomes
rather complicated by the presence of the non-local potentials.

In this paper, the Dirac negative energy spectrum or the single
anti-nucleon spectrum in atomic nucleus such as $^{16}$O  will be
investigated within the DDRHF theory in order to understand the
relativistic symmetry with non-local potentials. The corresponding
spin symmetry and its origin will be examined, in particular the
role of the Fock terms.

\section{Theoretical Framework}

The starting point of the DDRHF theory is an effective Lagrangian
density $\scr L$ \cite{Long:2006}, which contains the degrees of
freedom associated with the nucleon field ($\psi$), two isoscalar
meson fields ($\sigma$ and $\omega$), two isovector meson fields
($\pi$ and $\rho$) and the photon field ($A$). Then the effective
Hamiltonian $\scr H$ is obtained with the general Legendre
transformation. On the level of the mean field approximation, the
energy functional $\scr E$ is obtained by taking the expectation of
the Hamiltonian $\scr H$, where both the Hartree (direct) and Fock
(exchange) terms are kept. Finally, the Dirac equations, i.e. the
equations of motion of nucleons, are obtained via the variation of
the energy functional $\scr E$.

For spherical nuclei, the nucleon Dirac spinor can be written as,
\begin{equation}\label{Dirac spinor}
    f_\alpha(\bo r)=\ff r
    \lb \begin{array}{l}
        iG_{n_a}(r)\scr Y^{l_a}_{j_am_a}(\hat{\bo r})\\
        -F_{\tilde{n}_a}(r)\scr Y^{\tilde{l}_a}_{j_am_a}(\hat{\bo r})
    \end{array}\rb
    \chi_\ff2(\tau_a),
\end{equation}
where $\chi_\ff2(\tau_a)$ is the isospinor, $\scr Y^{l_a}_{j_am_a}$
is the spherical harmonics spinor and $\scr
Y^{\tilde{l}_a}_{j_am_a}(\hat{\bo r}) =-\hat{\bo\sigma}\cdot\hat{\bo
r}\scr Y^{l_a}_{j_am_a}(\hat{\bo r})$ with $\tilde{l}_a = 2j_a-l_a$.
For the negative energy states, the lower component $F(r)$ is
dominant. The states are labelled by $\{\tilde{n}\tilde{l}jm\}$ with
the relation
 \begin{equation}\label{nodes relation}
    n=\tilde{n},\quad\mbox{for\ }\kappa>0,
    \qquad
    n=\tilde{n}+1,\quad\mbox{for\ }\kappa<0,
\end{equation}
in analogy to Ref. \cite{Leviatan:2001}. The spin symmetry concerns
the near degeneracy of the states ($\tilde n, \tilde l, j=\tilde
l\pm 1/2 $). In the following equations, the sub-index will be
omitted for simplicity.

The radial Dirac equations are the coupled integro-differential ones
due to the non-local Fock terms $X$ and $Y$ \cite{Bouyssy:1987},
\begin{subequations}\label{RHF equations}
    \begin{equation}
        EG(r) = -\ls \frac{d}{dr}-\frac{\kappa}{r}\rs F(r)
            +\ls M+\Sigma_S(r)+\Sigma_0(r)\rs G(r) +Y(r),
    \end{equation}
    \begin{equation}
        EF(r) = +\ls \frac{d}{dr}+\frac{\kappa}{r}\rs G(r)
            -\ls M+\Sigma_S(r)-\Sigma_0(r)\rs F(r) +X(r).
    \end{equation}
\end{subequations}
Introducing the effective local potentials $X_G, X_F, Y_G$ and $Y_F$
by the definitions,
\begin{subequations}\label{localize}
    \begin{eqnarray}
        X(r) &=& \frac{G(r)X(r)}{G^2+F^2}G(r)+\frac{F(r)X(r)}{G^2+F^2}F(r)
            \equiv X_G(r)G(r)+X_F(r)F(r),
    \end{eqnarray}
    \begin{eqnarray}
        Y(r) &=& \frac{G(r)Y(r)}{G^2+F^2}G(r)+\frac{F(r)Y(r)}{G^2+F^2}F(r)
            \equiv Y_G(r)G(r)+Y_F(r)F(r),
    \end{eqnarray}
\end{subequations}
the integro-differential equations Eq. (\ref{RHF equations}) can be
formally rewritten as equivalent differential ones,
\begin{subequations}\label{localized RHF equations}
    \begin{equation}
        \ls \frac{d}{dr}-\frac{\kappa}{r}-Y_F(r)\rs F(r)
            -\ls V_+(r)-E\rs G(r)=0,
    \end{equation}
    \begin{equation}
        \ls \frac{d}{dr}+\frac{\kappa}{r}+X_G(r)\rs G(r)
            +\ls V_-(r)-E\rs F(r)=0,
    \end{equation}
\end{subequations}
where $V_+\equiv V_+^D+Y_G$, $V_-\equiv V^D+X_F$, and
\begin{equation}\label{local potentials}
    V_+^D\equiv M+\Sigma_S+\Sigma_0,
    \qquad
    V^D \equiv \Sigma_0-\Sigma_S-M.
\end{equation}
In the above expressions, $\Sigma_S$ represents the scalar potential
from the Hartree terms, $\Sigma_0$ is the time component of the
vector potential, which contains the contributions from the Hartree
terms and the rearrangement terms induced by the density-dependence
of the meson-nucleon couplings \cite{Long:2006}, and $X_G$, $X_F$,
$Y_G$, $Y_F$ are the effective local potentials from the Fock terms.
The equations Eq. (\ref{localized RHF equations}) then can be solved
self-consistently with the same numerical method as in RMF
\cite{Meng:1998a}.

From the radial Dirac equation Eq. (\ref{localized RHF equations}),
the Schr\"{o}dinger-type equation for the dominant component $F(r)$
can be obtained as,
\begin{equation}\label{Schrodinger type}
    \ff{V_+ - E}\Lb F^{''}+\lb V_1^D+V_1^E\rb F'
        +\ls V_{\mCB}+V^D_{\mSOP}+V^E_{\mSOP}
        \rs F\Rb+V^D F+V^E F
    = EF,
\end{equation}
where $V_{\mCB} = \frac{\kappa(1-\kappa)}{r^2}$ and $V_{\mSOP}$
correspond to the centrifugal barrier (CB) and spin-orbit potential
(SOP), respectively. In the above equation, the Hartree and Fock
terms for $V_1$, $V_{\mSOP}$ and $V$ read as
\begin{subequations}\begin{align}
    V_1^D=&-\frac{{V_+^D}'}{V_+-E}, &
    V_1^E=&X_G-Y_F-\frac{Y_G'}{V_+-E},\\
    V^D_{\mSOP}=&\frac{\kappa}{r}\frac{{V_+^D}'}{V_+-E}, &
    V^E_{\mSOP}=&\frac{\kappa}{r}\lb\frac{Y_G'}{V_+-E}-X_G-Y_F\rb,\\
    V^D=&\Sigma_0-\Sigma_S-M, &
    V^E=&X_F+\ff{V_+-E}\lb Y_F\frac{V_+'}{V_+-E}-Y'_F-X_GY_F\rb.
\end{align}\end{subequations}
One may note that the denominator ${V_+-E}$ contains a state
dependent potential $Y_G$. However, as the quantity $Y_G$ is around
a few MeV and is negligible in comparison with ${V_+-E}$ which is of
the order of 1 GeV, the Eq. (\ref{Schrodinger type}) is accurate
enough to estimate the Hartree and Fock contributions. Similar
argument also holds for the time component of the vector potential
$\Sigma_0$ which contains the rearrangement term from Fock channels.

\section{Results and Discussion}

Solving the DDRHF equations Eq. (\ref{localized RHF equations}) with
the parameter set PKO1 \cite{Long:2006} in coordinate space
self-consistently as in RMF \cite{Meng:1998a}, the neutron and
proton single-particle energies can be obtained. We take the nucleus
$^{16}$O as an example to examine  the negative energy spectrum and
its spin symmetry.

\begin{figure}[htbp]\centering
\includegraphics[width=0.45\textwidth]{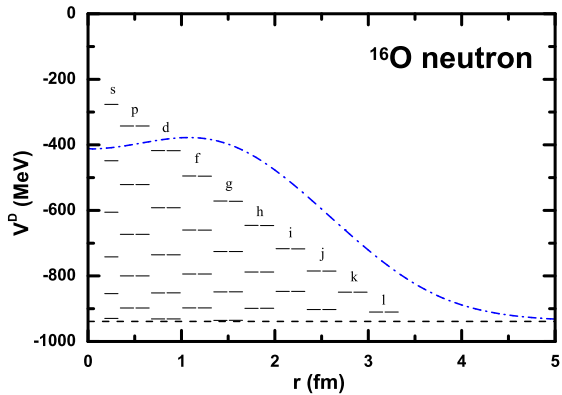}
\caption{(color online) Single neutron spectrum in the Dirac sea for
$^{16}$O calculated by DDRHF with PKO1. The dash-dot line represents
the Hartree potential $V^D$. For each pair of the spin doublets, the
left levels are those with $ j = \tilde{l}-1/2$ and the right ones
with $j = \tilde{l}+1/2$.} \label{fig:level}
\end{figure}

In Fig. \ref{fig:level}, all the bound single neutron states in the
Dirac sea for $^{16}$O are given. The dash-dot line represents the
corresponding Hartree potential $V^D$ which is not strong enough for
the $0s$ and $0p$ orbits, the importance and contribution of the
Fock terms is thus illustrated. For each pair of the spin doublets,
the left levels are those with $ j = \tilde{l}-1/2$ and the right
ones with $j = \tilde{l}+1/2$. It can be clearly seen that the spin
symmetry is well conserved in the Dirac sea.

\begin{figure}[htbp]\centering
\includegraphics[width=0.45\textwidth]{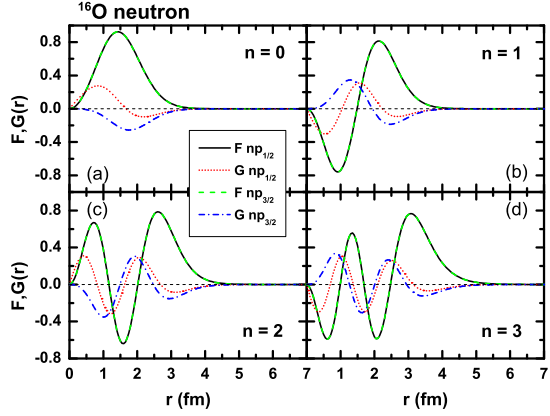}
\caption{(color online) Radial Dirac wave functions of the spin
doublets $p$ orbits in the negative energy spectrum of $^{16}$O
calculated by DDRHF with PKO1. Panels (a), (b), (c), and (d) are for
$0p$, $1p$, $2p$, and $3p$ spin doublets, respectively.}
\label{fig:wavep}
\end{figure}

Taking $p$ orbits with $n=0,1,2,3$ as examples, the Dirac wave
functions of spin partners are shown in Fig. \ref{fig:wavep}. The
dominant components $F(r)$ for the spin doublets are almost
identical, whereas the small components $G(r)$ show dramatic
deviations from each other due to the node relation given in Eq.
(\ref{nodes relation}). The features of the spin partners for both
the energies and wave functions are similar to those in RMF
\cite{Zhou:2003}. In the following, the origin and mechanisms of the
spin symmetry will be investigated in comparison with those in RMF.

%---------------------------splitting-------------------------------------------
%
\begin{figure}[htbp]\centering
\includegraphics[width=0.45\textwidth]{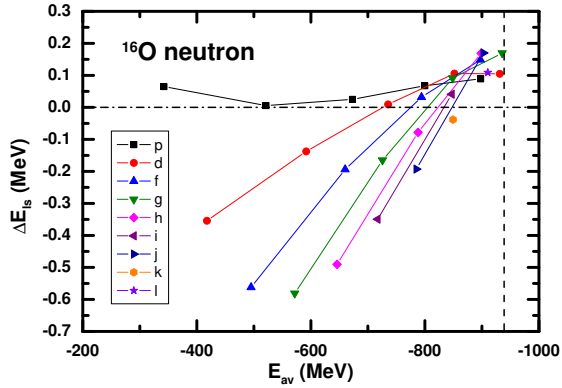}
\caption{(color online) Spin-orbit splitting $\Delta E_{\rm ls} =
E_{n\tilde{l}_{\tilde{l}+1/2}}-E_{n\tilde{l}_{\tilde{l}-1/2}}$ in
the negative energy spectrum of $^{16}$O versus the average binding
energy $E_{\rm av} =
(E_{n\tilde{l}_{\tilde{l}+1/2}}+E_{n\tilde{l}_{\tilde{l}-1/2}})/2 $
calculated by DDRHF with PKO1. The vertical dashed line shows the
continuum limit. }
    \label{fig:splitting}
\end{figure}

The spin-orbit splittings $\Delta E_{\rm ls} =
E_{n\tilde{l}_{\tilde{l}+1/2}}-E_{n\tilde{l}_{\tilde{l}-1/2}}$ in
the negative energy spectrum of $^{16}$O versus the average binding
energies $E_{\rm av} =
(E_{n\tilde{l}_{\tilde{l}+1/2}}+E_{n\tilde{l}_{\tilde{l}-1/2}})/2$
%calculated by DDRHF with PKO1
are given in Fig. \ref{fig:splitting}. In comparison with the RMF
results (see Fig. 2 in Ref. \cite{Zhou:2003}), the DDRHF results
have the following characteristics: 1) the spin-orbit splittings are
smaller; 2) the spin-orbit splittings fluctuate with $E_{\rm av}$,
in contrast with the monotonous decreasing in the RMF case,  when
approaching the continuum limit; 3) in RMF the spin-down state ($j =
\tilde l - 1/2$) is always lower than its spin-up partner ($j =
\tilde l + 1/2$), while in DDRHF this occurs only for the $p$ orbits
and states near the continuum limit.

%--------------------------Effective potential------------------------

In order to understand the origin of the spin symmetry in DDRHF and
the relative positions of the spin-up state and its spin-down
partner, the effective potentials $V$ in Eq. (\ref{Schrodinger
type}) as well as the relations between the centrifugal barrier and
the spin-orbit potential will be investigated.

\begin{figure}[htbp]\centering
\includegraphics[width=0.45\textwidth]{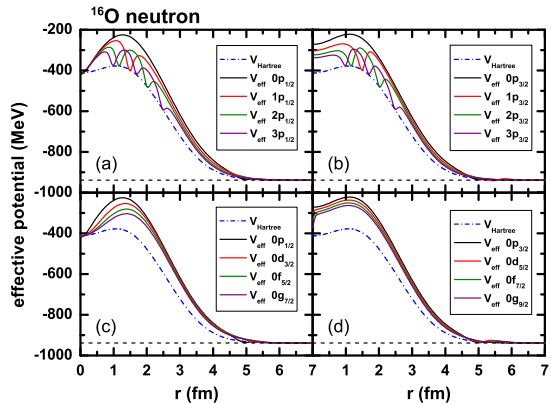}
\caption{(color online) Effective potentials in the negative energy
spectrum of $^{16}$O calculated by DDRHF with PKO1, (a) for states
$np_{1/2}$ with $n=0,1,2,3$, (b) for states $np_{3/2}$ with
$n=0,1,2,3$, (c) for states $0p_{1/2}$, $0d_{3/2}$, $0f_{5/2}$, and
$0g_{7/2}$, (d) for states $0p_{3/2}$, $0d_{5/2}$, $0f_{7/2}$, and
$0g_{9/2}$. The Hartree part is labelled with dash-dotted lines.}
    \label{fig:effpotl}
\end{figure}

The effective potentials $V$ for $p$, $d$, $f$, and $g$ states in
the negative energy spectrum of $^{16}$O calculated by DDRHF with
PKO1 are shown in Fig. \ref{fig:effpotl}, together with the Hartree
part $V^D$ (dash-dotted line). As seen in the Schr\"{o}dinger-type
equation Eq. (\ref{Schrodinger type}), the effective potential $V$
is composed of two parts, $V^D$ the Hartree potential from the
direct terms, and $V^E$ the equivalent local potential from the
exchange terms. The state dependence of the effective potential $V$
comes from the contribution of the exchange terms.

Corresponding to the nodes of the dominant component $F(r)$, there
exist fluctuations in the effective potentials $V$, which is brought
in by the localization of non-local terms $X$ and $Y$ in Eq.
(\ref{localize}). In addition, the contributions of Fock terms to
the effective potentials tend to be slightly weaker when $E_{\rm
av}$ approaches the continuum limit, or for larger orbital angular
momenta $\tilde{l}$.

Comparing the left and right panels of Fig. \ref{fig:effpotl}, it is
found that the effective potentials at $r=0$ are different between
the spin partner states. This is due to the different asymptotic
behaviors of the radial Dirac wave functions for spin doublets at
$r=0$,
\begin{equation}\label{wave function asymptotic}
    \begin{array}{ll}
        \displaystyle\lim_{r\rightarrow0}\frac{G(r)}{F(r)}\propto r, & \mbox{for}\quad \kappa>0,\\
        \displaystyle\lim_{r\rightarrow0}\frac{F(r)}{G(r)}\propto r, & \mbox{for}\quad \kappa<0.
    \end{array}
\end{equation}

Within the RMF framework, it has been pointed out that the strong
centrifugal barrier and weak spin-orbit potential lead to the
pseudo-spin symmetry in the single nucleon spectrum
\cite{Meng:1998PRCps} and the spin symmetry in the single
anti-nucleon spectrum \cite{Zhou:2003}.

\begin{figure*}[htbp]\centering
\includegraphics[width=0.80\textwidth]{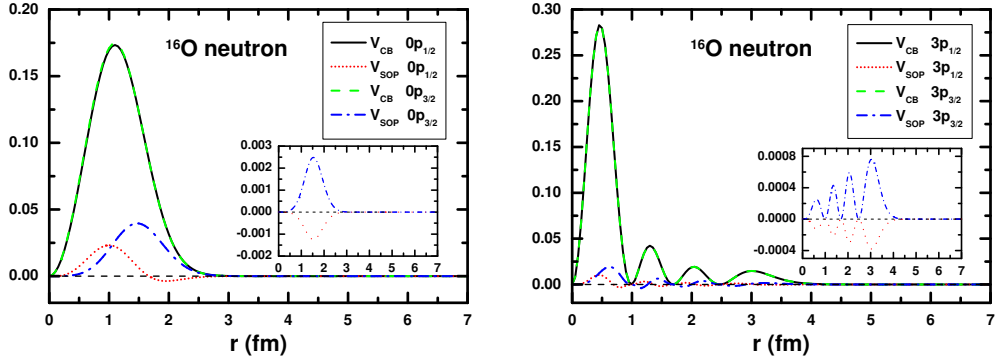}
\caption{(color online) Centrifugal barriers $V_{\mCB}$ and
spin-orbit potentials $V_{\mSOP}$ multiplied by the factor $\mp
F^2/(V_+-E)$ for the spin doublets ($\nu0p_{1/2},\nu0p_{3/2}$) (left
panel) and ($\nu3p_{1/2},\nu3p_{3/2}$) (right panel) in the negative
energy spectrum of $^{16}$O. The insets show the Hartree
contributions of the spin-orbit potentials. }
    \label{fig:cbsop03pD}
\end{figure*}

In Fig. \ref{fig:cbsop03pD} are shown the centrifugal barriers
$V_{\mCB}$ and the spin-orbital potentials $V_{\mSOP}$ multiplied by
the factor $\mp F^2/(V_+-E)$ for the spin doublets $0p$ and $3p$,
and their integrals over $r$ are respectively proportional to their
contributions to the single-particle energy. It is clearly shown
that the contribution of the centrifugal barriers $V_{\mCB}$ is much
larger than that of the spin-orbital potentials $V_{\mSOP}$.
Therefore, it can be concluded that similar reasons as in RMF lead
to the spin symmetry in the negative energy spectrum in DDRHF, and
the spin-orbit splitting is due to the different spin-orbit
potentials $V_{\mSOP}$ of the spin doublets.

In the insets of Fig. \ref{fig:cbsop03pD} are given the Hartree
contributions to the spin-orbit potentials. It is found that the
contributions from the Fock terms to $V_{\mSOP}$ are one order of
magnitude larger than those from the Hartree terms. Therefore, the
Fock terms must play important roles in the spin-orbit splitting of
the spin doublets.

%--------------------------contribution from each piece------------------

From Eq. (\ref{Schrodinger type}), the contributions to the
single-particle energies $E$ from different channels can be
estimated quantitatively. For example, the CB contribution can be
calculated by
\begin{equation}\label{CB contribution}
    \ff{ \displaystyle \int_0^\infty F^2dr}  \int_0^\infty  \frac{V_{\mCB}}{V_+-E}F^2dr.
\end{equation}
In Table \ref{Tab:contribution} are shown the contributions to the
single-particle energies and spin-orbit splittings for the spin
doublets $0p$ and $3p$. It is confirmed that the energy
contributions from $V_{\mCB}$ are much larger than those from
$V_{\mSOP}$ and the contribution from the Fock terms $V_{\mSOP}^E$
is dominant in $V_{\mSOP}$.

\begin{table*}[htbp]
   \centering
    \caption{The contributions from different channels (see Eq. (\ref{Schrodinger type}))
        to the single-particle energies $E$
        as well as the spin-orbit splittings $\Delta E$
        for the spin doublets ($\nu0p_{1/2},\nu0p_{3/2}$) and
        ($\nu3p_{1/2},\nu3p_{3/2}$) in the negative energy spectrum of $^{16}$O.
        The results are calculated by DDRHF with PKO1 and
        all units are in MeV.}\label{Tab:contribution}
    \begin{tabular}{crrrrrrrrr}\hline\hline
            state & \multicolumn{1}{c}{$F''$} & \multicolumn{1}{c}{$V_{\mCB}$} &
              \multicolumn{1}{c}{$V_1^D$} & \multicolumn{1}{c}{$V_{\mSOP}^D$} &
              \multicolumn{1}{c}{$V^D$} &
              \multicolumn{1}{c}{$V_1^E$} & \multicolumn{1}{c}{$V_{\mSOP}^E$} &
              \multicolumn{1}{c}{$V^E$} &
              \multicolumn{1}{c}{$E$}   \\  \hline
            $\nu 0p_{1/2}$  &   $-45.32$  &   $-41.92$   &   $0.21$  &  $-0.26$
                            &  $-416.00$  &     $5.35$   &   $3.65$  &  $151.87$  &    $-342.43$  \\
            $\nu 0p_{3/2}$  &   $-45.41$  &   $-42.19$   &   $0.21$  &  $0.53$
                            &  $-415.68$  &     $2.52$   &   $8.18$  &  $149.49$  &    $-342.37$
                            \\ \hline
            $\Delta E$      &    $-0.09$  &    $-0.27$   &   $0.00$  &  $0.79$
                            &     $0.32$  &    $-2.83$   &   $4.53$  &  $-2.38$   &       $0.06$  \\ \hline\hline
             $\nu 3p_{1/2}$ &  $-211.40$  &   $-38.44$   &   $0.01$  &  $-0.12$
                            &  $-611.42$  &     $4.77$   &   $0.90$  &  $55.82$   &    $-799.91$  \\
            $\nu 3p_{3/2}$  &  $-211.39$  &   $-38.42$   &   $0.01$  &  $0.24$
                            &  $-611.47$  &     $3.80$   &   $2.00$  &  $55.41$   &    $-799.84$
                            \\ \hline
            $\Delta E$      &     $0.01$  &     $0.02$   &   $0.00$  &  $0.36$
                            &    $-0.05$  &    $-0.97$   &   $1.10$  &  $-0.41$   &       $0.07$ \\
            \hline\hline
    \end{tabular}
\end{table*}

In Table \ref{Tab:contribution}, it is found that the contributions
from $V^D_{\mSOP}$, $V_1^E$, $V_{\mSOP}^E$ and $V^E$ to the
spin-orbit splitting are substantial. However, their contributions
are counteracted by one another to preserve the spin symmetry. This
kind of sophisticated cancellation implies that a weaker spin-orbit
potential $V_{\mSOP}$ does not mean better conserved spin symmetry,
as for $0p$ and $3p$ orbits.

\begin{figure}[htbp]\centering
\includegraphics[width=0.45\textwidth]{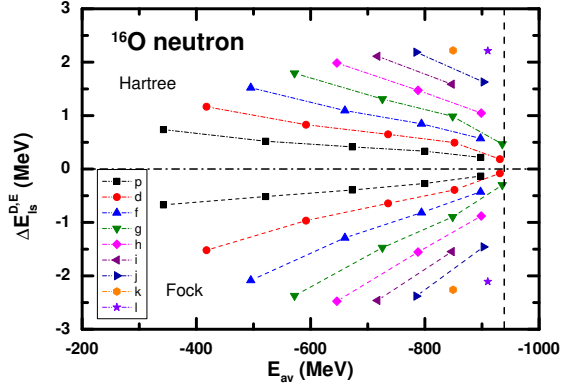}
\caption{(color online) Hartree and Fock contributions to spin-orbit
splitting
    in the negative energy spectrum of $^{16}$O
    versus the average energy of the spin doublets.
    The vertical dashed line shows the continuum limit.}
    \label{fig:splittingHF}
\end{figure}

To further confirm the role of the Fock terms, the contributions
from the Hartree and Fock channels to the spin-orbit splittings in
the negative energy spectrum of $^{16}$O versus the average energies
of the spin doublets are shown in Fig. \ref{fig:splittingHF}, where
the Fock part includes the contributions from the terms $V^E_1$,
$V_{\mSOP}^E$ and $V^E$ and the rest is gathered into the Hartree
part. It is found that the absolute contributions from both Hartree
and Fock parts decrease monotonously with the average energy $E_{\rm
av}$. The contributions from the Hartree terms have an energy
dependence similar to those in RMF \cite{Zhou:2003}. The
contributions from Fock terms have an opposite tendency and cancel
with the Hartree ones, thus leading to better spin symmetry. The
competition between the Hartree and Fock terms will determine the
sign of the spin-orbit splitting and this explains why the
spin-orbit splittings in DDRHF fluctuate with $E_{\rm av}$ in Fig.
\ref{fig:splitting} instead of monotonously decreasing as in the RMF
case.

In order to get a deeper understanding of the cancellation and
competition between the Hartree and Fock terms, we first separate
the different meson contributions to single-particle energies. It is
found that in both Hartree and Fock contributions, the isoscalar
mesons, $\sigma$ and $\omega$, play dominant roles in the spin-orbit
splitting, while the contributions from the $\rho$-, $\pi$-mesons,
and the rearrangement terms are negligible. Then, to make the
mathematical structure simple and clear, one could replace the
finite range Yukawa propagators with a pure delta function
$\delta(\bo r_1-\bo r_2)$ for these two heavy isoscalar mesons, but
keeping their Dirac scalar and vector couplings. In this simple
picture, it is found that the direct term of the $\sigma$-meson
makes the spin-orbit splitting positive, whereas that of the
$\omega$-meson makes the splitting negative. Since the attractive
$\sigma$ field is slightly stronger than the repulsive $\omega$
field in realistic nuclei, the net Hartree contribution to the
splitting is positive, as shown in the upper part of
Fig.~\ref{fig:splittingHF}. Meanwhile, it is also found analytically
that the effect of the $\sigma$ exchange term is roughly one half as
the effect of its direct term, but with an opposite sign. The effect
of the $\omega$ exchange term almost vanishes due to the
cancellation between the time and space components. Therefore, the
net Fock contribution to the splitting is negative and comparable to
the Hartree contribution, as shown in the lower part of
Fig.~\ref{fig:splittingHF}. All the above discussions for the case
of $^{16}$O are also valid for heavier nuclei, e.g., $^{208}$Pb.

%------------------------quenched----------------------------------------

\begin{figure}[htbp]\centering
\includegraphics[width=0.45\textwidth]{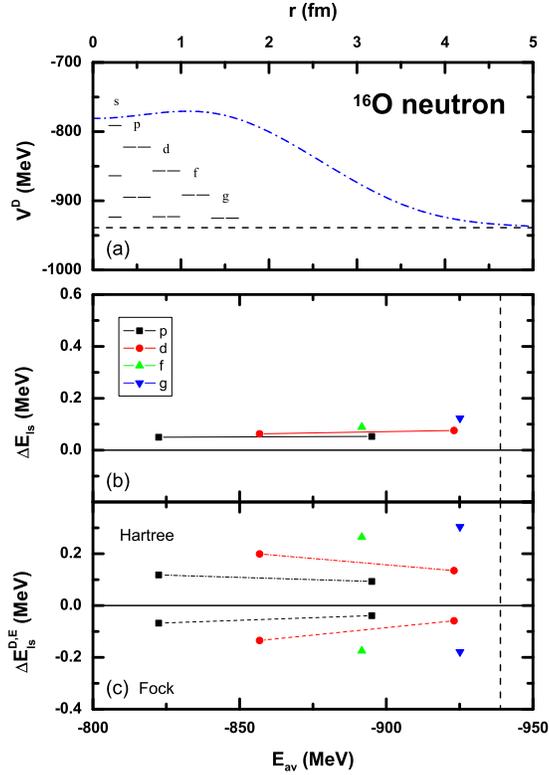}
\caption{(color online) Respectively same  as Fig.~\ref{fig:level}
    (a), Fig.~\ref{fig:splitting} (b), and Fig.~\ref{fig:splittingHF} (c),
    but with $g_{i\bar N\bar N} = g_{i\bar N N} = g_{iN\bar N}
        = 0.3 g_{iNN}$.}
    \label{fig:quench}
\end{figure}

It is known that the presence of strong annihilation channels and
various many-body effects could cause significant deviations from
the G-parity values of the meson-antinucleon couplings
\cite{Mishustin:2005} and a global fits to antiprotonic X-rays and
radiochemical data indicates the need to use reduction factors
\cite{Friedman:2005}.  Therefore in order to exam whether the spin
symmetry still persist in the face of these effects, the Dirac
negative energy spectrum is calculated with the coupling constants
$g_{i\bar N\bar N} = g_{i\bar N N} = g_{iN\bar N} = 0.3 g_{iNN}$,
where $i=\sigma, \omega, \rho$. In Fig. 7, all the bound single
neutron states in the Dirac sea thus obtained are shown, together
with the spin-orbit splitting as well as its Hartree and Fock
contributions. It is found that the spin symmetry is still well
conserved even in the case where significant deviations from the
G-parity values of the meson-antinucleon couplings due to the strong
annihilation channels and various many-body effects occur.

%------------------------------------------------------------------------
\section{Summary}

In summary, the spin symmetry in the negative energy spectrum and
its mechanism are investigated within the DDRHF theory by taking the
nucleus $^{16}$O as an example.

Similarly to RMF, the spin symmetry in the negative energy spectrum
is found to be a good approximation and the dominant components
$F(r)$ of the Dirac wave functions for the spin doublets are nearly
identical, as the centrifugal barrier is much stronger than the
spin-orbit potential.

However, it is found that the Fock terms are dominant in the
spin-orbit potential, which induce the state dependence of the
effective potential and play essential roles in spin-orbit
splitting.

Classifying the contributions to the spin-orbit splitting into the
Hartree and Fock parts, it is found that the Hartree part has an
energy dependence similar to those in RMF \cite{Zhou:2003}, while
the Fock part has an opposite tendency and cancels with the Hartree
part, thus leading to good spin symmetry.

The competition between the Hartree and Fock terms determines the
subtle spin-orbit splitting, which explains the fluctuation of the
spin-orbit splittings in DDRHF instead of monotonously decreasing
with $E_{\rm av}$  as in the RMF case.

\section*{Acknowledgments}
This work is partly supported by State 973 Program 2007CB815000, the
NSF of China under Grant Nos. 10435010, 10775004 and 10221003, and
the CNRS(France) - NSFC(China) PICS program no. 3473. One of the authors (H.L.) is
grateful to the French Embassy in Beijing for the financial support
for his stay in France.

%=====================================================================================================

\end{document}